# Exploiting synthetic lethal vulnerabilities for cancer therapy

Sriganesh Srihari

Institute for Molecular Bioscience, The University of Queensland, St. Lucia, Queensland 4072, Australia.

**Abstract**

*Synthetic lethality* refers to a combination of two or more genetic events (typically affecting different genes) in which the co-occurrence of the events results in cell or organismal lethality, but the cell or organism remains viable when only one of the events occurs. Synthetic lethality has gained attention in the last few years for its value in selective killing of cancer cells: by targeting the synthetic lethal partner of an altered gene in cancer, only the cancer cells can be killed while sparing normal cells. In a recent study, we showed that mutual exclusive combinations of genetic events in cancer hint at naturally occurring synthetic lethal combinations, and therefore by systematically mining for these combinations we can identify novel therapeutic targets for cancer. Based on this, we had identified a list of 718 genes that are mutually exclusive to six DNA-damage response genes in cancer. Here, we extend these results to identify a subset of 43 genes whose over-expression correlates with significantly poor survival in estrogen receptor-negative breast cancers, and thus provide a promising list of potential therapeutic targets and/or biomarkers.

**Background**

Inhibiting activated oncogenes has been the primary approach for targeted cancer therapeutics. However, identifying targetable oncogenes might not be feasible in many tumours — for example, in tumours for which a distinctive causal oncogene cannot be found (*e.g.* triple-negative breast cancers, TNBCs [12]), or in tumours that are primarily driven by inactivation of (tumour suppressor) genes (*e.g.* BRCA-deficient tumours). Alternative means and ways of identifying targetable genes in these cancers remain to be explored.

Cancer cells are fundamentally (genetically) different from normal cells, and therefore even if cancer cells depend on the same cellular processes to maintain their viability and proliferation, the *manner* in which the cells utilize these processes to their benefit is fundamentally different from that of normal cells. Cancer cells often capitalize on 'alternative routes' and 'back up or by pass' pathways which normal cells do not activate in their normal course. These 'survival tricks' are picked up by cancer cells through selective pressure wherein the constantly

(genetically) evolving cancer cell mass gives rise to clones that gather the right combination of genetic alterations to enable them meander their ways to survival. However, in this constant pursuit for survival, cancer cells often acquire *vulnerabilities* which normal cells do not, and which if targeted can bring about the catastrophic death of cancer cells.

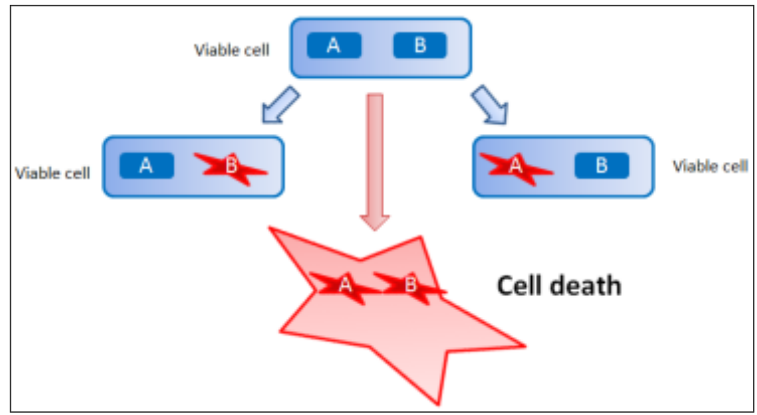

**Figure 1:** The concept of synthetic lethality.

One such vulnerability arises *via synthetic lethality* between the genetic alterations acquired by cancer cells. Synthetic lethality (SL) was first defined by Calvin Bridges in 1922 [1], and refers to the combination between two genetic alterations in which the co-occurrence of both alterations is lethal to cells, although cells remain viable when only one of the alterations occurs (**Figure 1**). SL has gained considerable attention over the last few years due to its value in understanding the essentiality of genes or their combinations — for example, individual and pairs of genes have been systematically knocked out in yeast to understand gene essentiality [2]. More recently, SL has gained attention as a therapeutic strategy to exploit genetic vulnerabilities acquired by cancer cells, and thereby to enable selective targeting of these cells [3,4]: by targeting the SL partner of a genetic alteration in cancer cells, these cells can be selectively killed while sparing (normal) cells which do not harbour that alteration.

The first breakthrough in SL-based cancer therapy came when a study [4], albeit as an initial proof of concept, showed that inhibition of poly (ADP-ribose) polymerase (PARP) enzymes in cancer cells that harbour inactivation (loss-of-function) events in the breast-cancer susceptibility genes *BRCA1* and *BRCA2* is dramatically lethal to these cells, although inhibition of PARP by itself is not lethal in normal cells. The first (Phase I) set of clinical trials (ClinicalTrials.gov number NCT00516373) based on inhibiting PARP using olaparib in sixty patients carrying *BRCA1* or *BRCA2* mutations showed dramatic remission in these patients without severe side effects unlike conventional chemotherapy, thus providing the first clinical evidence for potent SL-based cancer therapy [5].

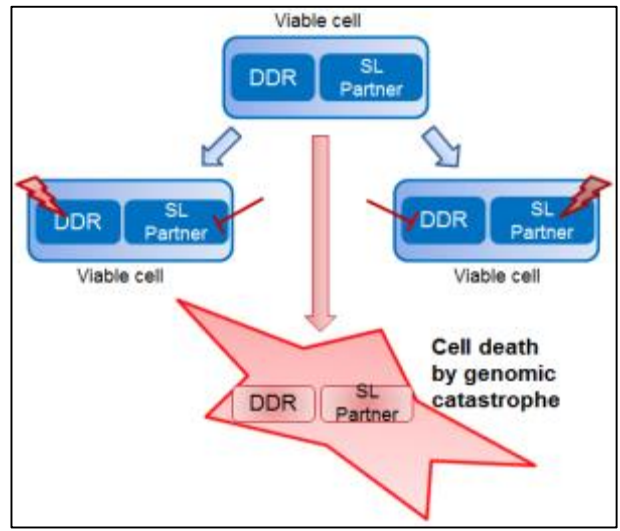

**Figure 2:** Genomic catastrophe induced by targeting the synthetic lethal (SL) partner of a DNA-damage response (DDR) gene.

Although still not completely understood, the mechanisms underlying BRCA-PARP synthetic lethality is explained using primarily three models [6, 7]. Cancer cells as a consequence of their accelerated cell-division cycles are exposed to considerable replication stress and they acquire significant genomic instability arising from frequent replication-fork stalling and the resultant DNA double-strand breaks (DSBs) during the DNA replication phase. Maintaining a stable genome is critical for cell survival, and the BRCA-mediated homologous recombination (HR) pathway is the primary pathway responsible for error-free repair of DSBs arising during replication. PARP along with other proteins (*e.g.* BLM and WRN) aid in restarting stalled replication forks and in converting collapsed replication forks arising during replication into DSBs for subsequent repair by HR. Consequently, the inhibition of PARP in cells already deficient in BRCA or certain other HR proteins (a condition referred to as "BRCAness") results in large number of stalled forks and the accumulation of lethal levels of DSBs which cannot be repaired, thus dealing a 'double blow' to the survivability of these cells — a condition called *genomic catastrophe*. An alternative model suggests that inhibition of PARP activates the non-homologous end joining (NHEJ) pathway to repair DSBs. Unlike HR which uses homologous strands as templates for the repair of DSBs, NHEJ operates by "cut and stitch" causing considerable genomic instability due to erroneous repair, and thereby a genomic catastrophe in cancer cells. The third model suggests the DNA-damaging role of chemical PARP inhibitors itself, wherein the inhibitor compound binds to the DNA causing an adduct that stalls DNA replication, as the reason for DNA damage and cell death. Irrespective of the specific underlying mechanism, the lethality observed in BRCA-deficient cells upon inhibition of PARP highlights a fundamental vulnerability that can be exploited for killing these cells.

The challenge therefore remains to identify cancer cell-specific SL vulnerabilities that can be similarly exploited for cancer therapy. [7] and [8] give a detailed and up-to-date account of the developments in this area; here, we describe the findings from a recent computational approach published in *Biology Direct* [9]. The authors in this work draw parallels between SL vulnerabilities and certain *mutual exclusive combinations* of genetic events that occur naturally in cancers. Mutual exclusivity (ME) refers to the observed lack of co-occurrence of any two genetic alterations in a large collection of cancers although each alteration by itself occurs in substantial subsets of the cancers. Although hinted in a few earlier works (*e.g.* [10]), the authors in [9] make a fairly formal (mathematical) argument to relate SL with ME combinations: if the two alterations are frequently observed in large subsets of cancers yet these two never or rarely co-occur then possibly their co-occurrence affects cancer cell viability (therefore, cancers harbouring their co-occurrence are rarely observed), and therefore many ME combinations could in fact be SL combinations occurring naturally in cancers.

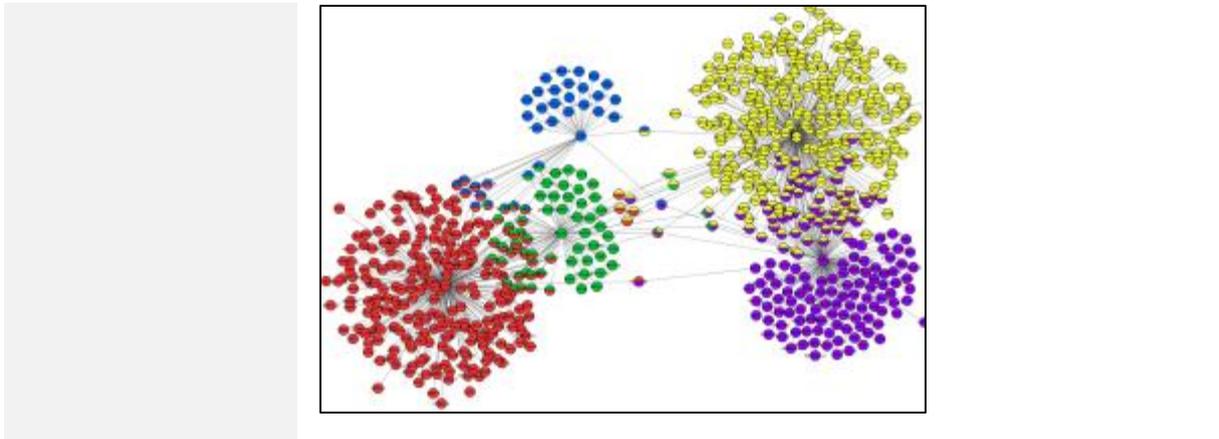

**Figure 3:** A network of 718 genes mutually exclusive (predicted to be synthetic lethal) to defects in at least one of the six DDR genes in the four cancers analysed in the study [9].

## Results

As a proof for their concept, the authors identified ME partner genes of six DNA-damage response (DDR) genes *ATM*, *BRCA1*, *BRCA2*, *CDH1*, *PTEN* and *TP53* by computationally screening for statistically significant ME combinations across four cancer types breast, ovarian, prostrate and uterine, using genomic and gene-expression data from The Cancer Genome Atlas. The expectation is that any further alterations in the partner genes of these defective DDR genes induces cancer cell death *via* a genomic catastrophe arising from accumulation of DNA damage (**Figure 2**). Based on this computational screen, about 700 genes were identified to be statistically significant for ME with at least one of the six DDR genes (**Figure 3**). Of these, the expression levels of a selected subset of 43 genes could separate out good and bad prognosis patients in an independent cohort of estrogen-receptor (ER)-negative breast cancer (which includes TNBCs) patients (**Figure 4**).

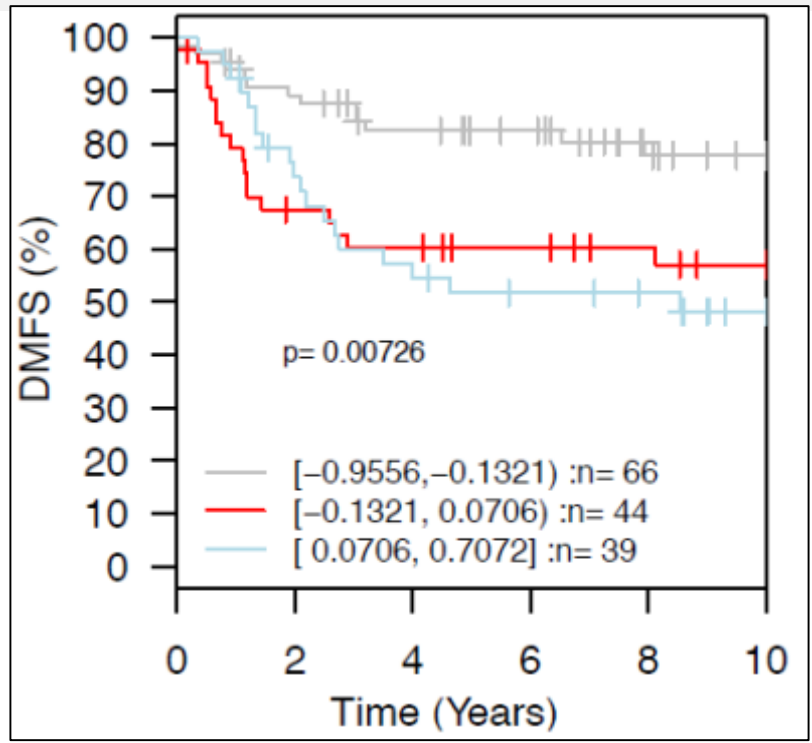

**Figure 4:** The 43 shortlisted genes from the ~700 mutual exclusive partners of DDR genes could divide estrogen-receptor (ER)-negative breast cancer patients from an independent cohort (GOBO) into good and bad prognosis groups. Higher expression of these 43 genes correlated with poor distant metastasis-free survival (DMFS) in these patients, thus indicating that inhibiting these genes could be a way to counter these cancers [blue = high expression; grey = low expression].

Interestingly, many of the ME partners identified in the study [9] turned out to be oncogenes that are overexpressed mainly in cancers that do not harbour any DDR alterations, whereas are never or rarely overexpressed in cancers with DDR alterations. This highlights an important mode of survival for cancer cells: these cells tend to maintain an *optimal* level of expression for oncogenes — neither too high (that leads to a genomic catastrophe) nor too low (that curtails their growth and proliferation) — in the context of a DDR defect to maintain their survivability. These oncogenes do not get picked up by routine scans for statistically significant (overexpressed) gene lists in cancers where the genes are expressed at optimally low levels.

However, these genes expose inherent *context-dependent vulnerabilities* in cancer cells which can be exploited for selective cancer-cell targeting. Here, if these ME partners are targeted in *conjunction* with alterations in DDR genes, cancer cells could be selectively killed while sparing the (normal) cells that do not harbour the DDR alterations. Indeed, in [11] the authors show that siRNA-mediated knockdown of *BRF2*, a mutual-exclusive partner of *BRCA1* and *BRCA2*, significantly reduced the viability of estrogen receptor-negative (ER-negative) cells, and in particular ER-negative/HER2-enriched MDA-MB-453 cells (**Figure 5**), but had relative much less impact on normal MCF10A cells. Therefore, the study [9] presents an interesting approach to identify targetable genes in cancers that do not harbour a distinctive causal oncogene or are primarily driven by inactivation of tumour suppressor genes.

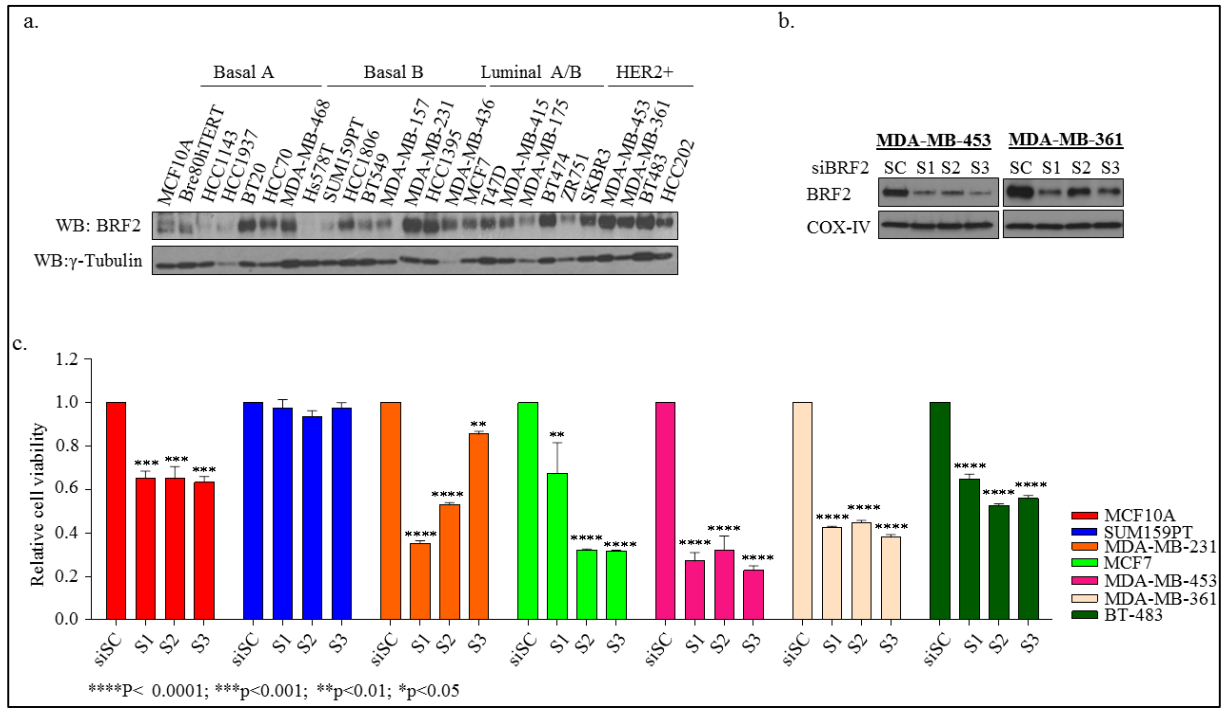

**Figure 5:** Effect siRNA-mediated knockdown of BRF2 in a panel of breast normal and cancer cell lines. (a) Protein expression levels of BRF2 measured using Western blotting. (b) Knockdown efficiencies of siRNAs against BRF2. (c) Cell viability upon BRF2 knockdown.